\documentclass[aps,pra,reprint,10pt,a4paper]{revtex4-1}
\usepackage[utf8]{inputenc}
\usepackage[sc,osf]{mathpazo}
\usepackage{amsmath}
\usepackage[T1]{fontenc}
\usepackage{latexsym}
\usepackage{amssymb}
\usepackage[colorlinks=true,citecolor=blue,urlcolor=blue]{hyperref}
\usepackage{color}
\usepackage{graphics,epstopdf}
\usepackage{soul}
\usepackage[demo]{graphicx}
\usepackage{capt-of}
\usepackage{lipsum}
\usepackage{adjustbox}
\usepackage[normalem]{ulem}
\usepackage[table,xcdraw]{xcolor}
\usepackage{braket}
\usepackage{physics}
\usepackage{ragged2e}

\usepackage{comment}

\begin{document}

\title{Nonclassical  resource for  continuous variable telecloning with non-Gaussian advantage\\
}

\author{Sudipta Das$^1$, Rivu Gupta$^2$, Himadri Shekhar Dhar$^{1,3}$, Aditi Sen(De)$^2$}

\affiliation{$^1$ Department of Physics, Indian Institute of Technology Bombay, Powai, Mumbai, Maharashtra - 400076, India \\
$^2$ Harish-Chandra Research Institute,  A CI of Homi Bhabha National Institute, Chhatnag Road, Jhunsi, Prayagraj - 211019, India \\
$^3$ Centre of Excellence in Quantum Information, Computation, Science and Technology, Indian Institute of Technology Bombay, Mumbai - 400076, India}

\begin{abstract}


The telecloning protocol distributes quantum states from a single sender to multiple receivers via a shared entangled state by exploiting the notions of teleportation and approximate cloning. 
We investigate the optimal telecloning fidelities obtained using both Gaussian and non-Gaussian shared resources. When the shared  non-Gaussian state is created by subtracting photons from both the modes of the Gaussian two-mode squeezed vacuum state, we demonstrate that higher telecloning fidelities can be achieved in comparison with its Gaussian counterpart. To quantify this advantage, we introduce a quadrature-based nonclassicality 
measure, which is capable of estimating the fidelity of the clones, both with Gaussian and non-Gaussian resource states. We further provide a linear optical setup for asymmetric telecloning of continuous variable states using a multimode entangled state.

%

\end{abstract}

\maketitle
\section{Introduction}
\label{sec:intro}

The laws of quantum mechanics enable information transmission protocols such as
dense coding~\cite{Bennett_PRL_1992}, teleportation~\cite{Bennett_PRL_1993}, secure key distribution~\cite{Gisin_RMP_2002}, and state transfer~\cite{Bose_PRL_2003}, that can outperform classical communication schemes. 
On the other hand, the same principles impose constraints on the physical processes~\cite{Landauer_PT_1991, Landauer_PA_1999}, leading to no-go theorems, which include no-cloning~\cite{Park_FP_1970, Wootters_Nature_1982, Dieks_PLA_1982, Yuen_PLA_1986}, no-broadcasting~\cite{Barnum_PRL_1996, Barnum_PRL_2007, Piani_PRL_2008}, and no-deletion theorems~\cite{Pati_Nature_2000} (also see~\cite{Pati_PRL_2007, Modi_PRL_2018, Nielsen_PRL_1997}). 
For instance, 
the no-cloning theorem~\cite{Wootters_Nature_1982}
prohibits the production of exact copies of an arbitrary quantum state and at the same time, is responsible for detecting eavesdroppers in quantum key distribution (QKD)~\cite{Gisin_RMP_2002}. Furthermore, approximate cloning strategies~\cite{Buzek_PRA_1996, Gisin_PRL_1997, Werner_PRA_1998, Scarani_RMP_2005} have been devised both universally~\cite{Bruss_PRL_1998} as well as in a state-dependent manner~\cite{Bruss_PRA_1998}, which provide bounds on the security of QKD (see~\cite{Linares_Science_2002, Cummins_PRL_2002, Martini_Nature_2002, Chen_PRA_2007, Yang_CTP_2007, Sabuncu_PRL_2007} for experimental realizations of quantum cloning).

Beyond point-to-point communication, the quantum telecloning protocol~\cite{Murao_PRA_1999, Dur_PRA_2001} leverages the inherent properties of both quantum teleportation and the no-cloning theorem to provide a resource-efficient method for distributing quantum information symmetrically among multiple parties. The protocol exhibits the importance of multipartite entanglement in communication networks. Successful demonstrations of telecloning schemes in laboratories have been reported with
photonic platforms~\cite{Zhao_PRL_2005, Koike_PRL_2006, Chiuri_PRL_2012}, NISQ computers~\cite{Pelofske_IEEE_2022}, and superconducting processors~\cite{pelofske_arxiv_2023}.
Further, a variation of telecloning, known as asymmetric telecloning~\cite{Cerf_PRL_2000, Cerf_JMO_2000, Murao_PRA_2000, Ghiu_PRA_2003, Zhao_PRL_2005} has also been proposed, in which different receivers obtain clones with different accuracies, thereby making it crucial for secure quantum communication.

Going beyond discrete variables, continuous variable (CV) systems~\cite{Gerry_2004, Serafini_2017}, characterized by quadrature operators with infinite spectrum, comprise physical systems that are essential for the experimental realization of information-theoretic protocols~\cite{Reck_PRL_1994, Furusawa_Science_1998, Lloyd_PRL_1999, Calsamiglia_APB_2001, Gottesman_PRA_2001, Fiurasek_PRL_2001, Li_PRL_2002, Schori_PRL_2002, Grosshans_Nature_2003, Bowen_PRA_2003, Liu_PRL_2021}. In this regime, the protocol of telecloning was introduced using a multimode Gaussian entangled state, distributed between a single sender and several receivers~\cite{van-Loock_PRL_2001}, allowing the production of clones with the same fidelity~\cite{Nielsen_2012}. Moreover, a reversible telecloning scheme~\cite{Zhang_PRA_2006}, constituting the preparation of a local clone (also known as anticlone), has been designed along with the proposal for telecloning with the aid of phase-conjugate input states~\cite{Zhang_PRA_2008}. 
On the experimental front, telecloning has also been exhibited using CV systems~\cite{Wang_PRA_2021}.


\begin{figure*}[ht]
\includegraphics[width=0.6\linewidth]{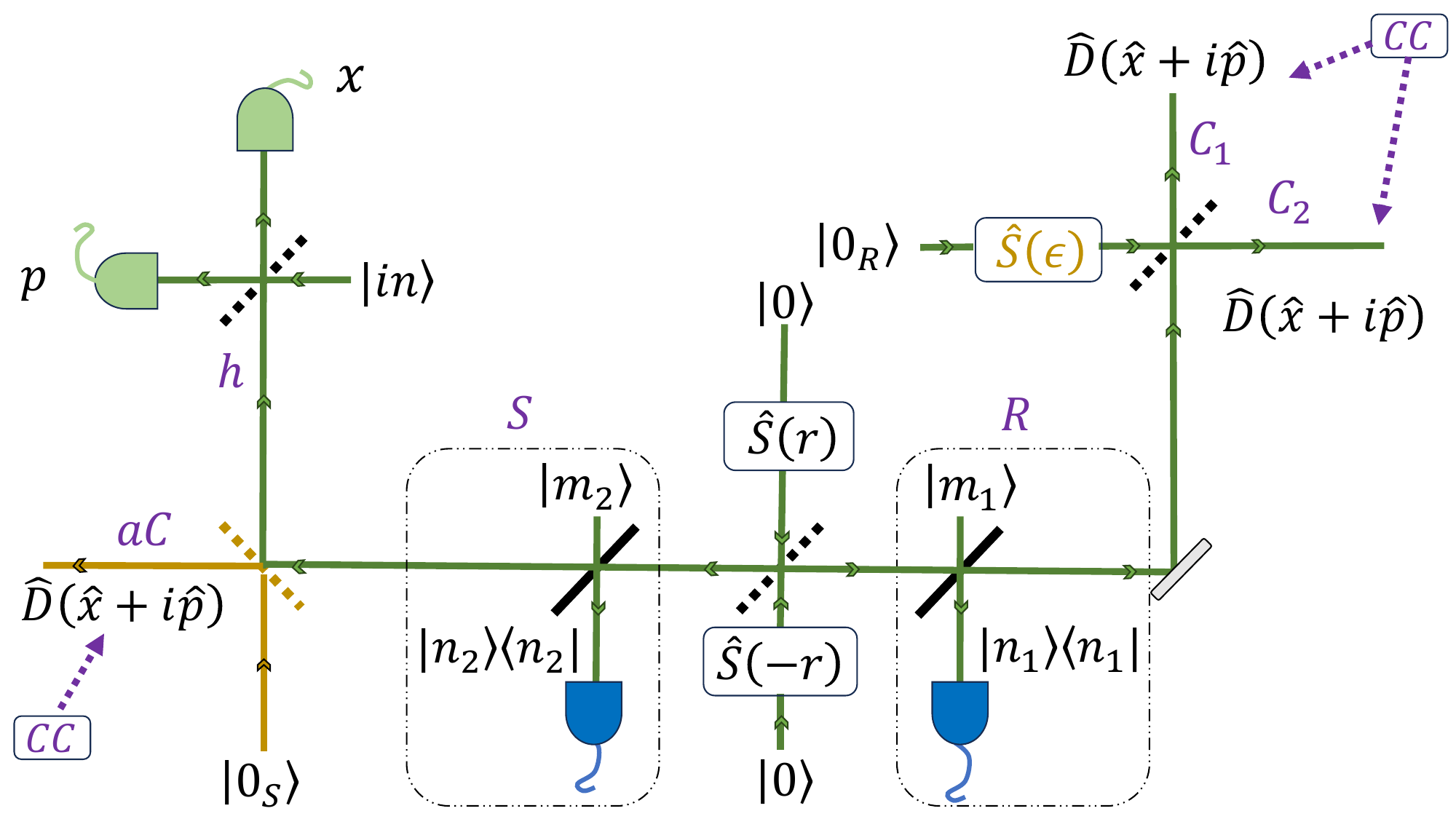}
\caption{\textbf{Schematic setup of telecloning for creating two clones.} Initially two vacuum modes, $\ket{0}$, one position squeezed $(\hat{S}(r))$ and the other momentum squeezed $(\hat{S}(-r))$ are passed through a balanced beam splitter to create the TMSV state. To de-Gaussify the modes, they are impinged on beam splitters of perfect transmissivity along with $m_i$ photon states followed by a heralded detection of $n_i$ photons. To create the $l$ photon-added state, $m_i = l$ and $n_i = 0$, whereas for the $l$ photon-subtracted state, $m_i = 0$ and $n_i = l$~\cite{Kumar_PRA_2022}. For the \textit{irreversible} protocol, one of the modes, $\mathcal{S}$, is combined with the input state, $\ket{\text{in}}$, at a balanced beam splitter for homodyne detection, $h$, of the quadratures $x$ and $p$. (For the \textit{reversible} protocol, the mode is initially combined with vacuum, $ v\mathcal{S} \equiv \ket{0_{\mathcal{S}}}$, at a balanced beam splitter. One of the two output modes is used for homodyne detection, whereas the other is used for the creation of an anticlone, $aC$.) The remaining resource mode, $\mathcal{R}$, is split into two using a balanced beam splitter and vacuum $\ket{0_{\mathcal{R}}}$ (for telecloning of squeezed input states, $\ket{0_{\mathcal{R}}}$ is further squeezed by $\hat{S}(\epsilon)$). The resulting modes are used for the production of the two clones, $C_1$ and $C_2$. The homodyne outcomes are classically communicated (CC) to the clone and anticlone modes whereafter a corresponding displacement $\hat{D}$ is performed. In the figure, all the tilted dashed lines represent $50/50$ (balanced) beam splitters, whereas the solid tilted lines are beam splitters of transmissivity unity. The blank rectangle at the receiver station is a mirror.}
\label{fig:sch} 
\end{figure*}

One of the central questions concerning any information processing task is -- ``\textit{what particular resource drives the quantum advantage evinced by the state applied for the protocol?}'' Since telecloning inherits the framework of teleportation, it is tempting to assume that entanglement~\cite{Horodecki_RMP_2009} is the key resource governing the performance of the protocol, although such correspondence has not been reported yet. By studying the maximum fidelity of telecloning with the entanglement of the shared resource state, we find that in the CV regime, the connection between the two is not straightforward. In particular, we study whether bimodal or genuine multimode entanglement is required to successfully produce distant clones in a CV photonic network. We establish that although entanglement is certainly necessary for any telecloning protocol to function, the fidelity of the clone received by any receiver station actually depends on other forms of non-classicality \cite{Duan_PRL_2000, Gehrke_PRA_2012, Yadin_PRX_2018} present between the sender and the corresponding receiver modes. To this end, we define an operational non-classicality measure, whose behavior is shown to be closely related to the fidelity of the clones.

From the perspective of quantum correlation content, it is known that non-Gaussian states~\cite{Walschaers_PRX_2021} possess higher entanglement~\cite{Navarrete_PRA_2012, Roy_PRA_2018, Fan_PRA_2018} than their Gaussian counterparts. Such states have been shown to outperform Gaussian states in various CV quantum protocols, such as quantum illumination~\cite{Fan_PRA_2018}, quantum teleportation~\cite{DellAnno_PRA_2007, DellAnno_PRA_2010, DellAnno_PRA_2010_2, Dhar_JPB_2015, Patra_PRA_2022, Chandan_PRA_2023-tele}, cryptography~\cite{Chen_OE_2023}, and steering~\cite{Olsen_PRA_2013}. Therefore, one can expect that the performance of telecloning can be improved by introducing non-Gaussianity in the protocol. We report that although non-Gaussian states lead to an increased fidelity, it is not universal.
Specifically, we show that non-Gaussian states, generated by the subtraction of photons from both the modes of a two-mode squeezed-vacuum (TMSV) state, can offer an advantage in telecloning over their parent Gaussian state. \textcolor{black}{Note that photon-addition and subtraction have been experimentally achieved~\cite{Dakna_EPJD_1998, Parigi_Science_2007, Ourjoumtsev_PRL_2007, Takahashi_NP_2010, Kurochkin_PRL_2014, Loaiza_NPJ_2019, Zhang_PRA_2020, Thapliyal_PRR_2024} and such states have been shown to possess enhanced capabilities of realizing information-processing tasks~\cite{Navarrete_PRA_2012, Kumar_PRA_2022} as compared to their Gaussian counterparts. Furthermore, most of these experiments are based on linear optics, and thus the generation of the resource state can be readily integrated into the telecloning protocol as shown in Fig. \ref{fig:sch}.} We further demonstrate that the non-classicality measure provides an estimate of the squeezing strength of the resource, at which the telecloning fidelity crosses the classical threshold both in the case of Gaussian and non-Gaussian states, \textcolor{black}{i.e., the usefulness of the resource states for telecloning can be readily verified using the nonclassicality measure, which can be determined using homodyne detection in the proposed scheme.} To our knowledge, such a connection has not been established in CV telecloning
Furthermore, we provide a setup for asymmetric telecloning in the CV regime through the generation of a genuinely entangled multimode resource state using only linear optical elements.

Our article is organized in the following way. First, we demonstrate that entanglement is not the only resource that governs the telecloning process in Sec.~\ref{sec:moti}. We then define our measure of non-classicality in Sec.~\ref{sec:non-cl} and illustrate how it properly explains whether a given resource state is suitable for telecloning CV states. In Sec.~\ref{sec:n-G_tele}, we study the telecloning of input coherent and squeezed states with the help of non-Gaussian photon-added and photon-subtracted states. We also discuss how the telecloning network needs to be modified when squeezed states need to be telecloned to various receivers. Sec.~\ref{sec:asymm_tele} lays down the asymmetric telecloning network comprising the framework for generating the multimode resource state. We conclude with discussions in Sec.~\ref{sec:conclu}.

\section{Quadrature variance as a telecloning resource}
\label{sec:moti}

Entanglement is a necessary ingredient for the successful implementation of several quantum information-theoretic tasks. For example, it is well known that bipartite entanglement serves as the key resource for dense coding~\cite{Bennett_PRL_1992} and teleportation~\cite{Bennett_PRL_1993}, whereas multiparty entanglement is necessary for measurement-based quantum computation~\cite{Briegel_NP_2009} and state transfer~\cite{Bose_PRL_2003}. However, for CV telecloning the role of entanglement is not so straightforward. In general, there are two primary telecloning protocols.

Let us first briefly describe the two telecloning protocols, which we refer to as the \textit{reversible protocol}, and the \textit{irreversible protocol}, respectively. The two aforementioned schemes differ in whether an anticlone is created \cite{Zhang_PRA_2006} or not \cite{van-Loock_PRL_2001}. While the clones produced do not contain the entire information of the original input state, the first protocol allows for the reconstruction of the same owing to the additional information present in the anticlone. Note that, in our work, we consider the production of only two clones in both protocols $(1 \to 2)$, an extension to a higher number of clones $(1 \to M)$ being straightforward. A schematic representation of the telecloning setup is provided in Fig.~\ref{fig:sch}.

To create the $1 \to 2$ telecloning network, one mode, $\mathcal{S}$, of a two-mode resource state, is given to the sender whereas the other mode, $\mathcal{R}$, is distributed among the two receivers by splitting it using a balanced beam splitter and vacuum, $v\mathcal{R} \equiv \ket{0_{\mathcal{R}}}$. For both protocols under study, the two receiver modes are $C_{1(2)} = \frac{\mathcal{R} \pm v\mathcal{R}}{\sqrt{2}}$. In the irreversible protocol, the mode at the sender's station is combined with the input state, followed by a homodyne measurement of the position and momentum quadratures. The outcomes are then communicated classically to the two distant receivers. We represent the mode at the homodyne station as $h$, i.e., $h = \mathcal{S}$ in the irreversible protocol. On the other hand, the reversible protocol entails splitting the sender's mode further into two modes using a balanced beam splitter and the vacuum $(v\mathcal{S} \equiv \ket{0_{\mathcal{S}}})$ -- one of these modes is then used for homodyne detection ($h = \frac{\mathcal{S} + v}{\sqrt{2}}$), whereas the remaining mode is used to create the anticlone. The success of the protocol can be measured in terms of the fidelity of the clone state and the input state, in phase space. Following the treatment elucidated in Ref.~\cite{Zhang_PRA_2006}, we provide the quadrature expressions for the two clones and one anticlone in the $1 \to 2$ telecloning arrangement. Note that, straightforward yet tedious calculations can be used to easily derive the same for the $1 \to M$ telecloning protocol.

\noindent \textbf{Irreversible protocol: }
\begin{eqnarray}
\nonumber \hat{x}_{C} = \frac{\hat{x}_\mathcal{R}}{\sqrt{2}} + \hat{x}_{\text{in}} - \hat{x}_\mathcal{S} - \frac{\hat{x}_{v\mathcal{R}}}{\sqrt{2}}, \\
\hat{p}_{C} = \frac{\hat{p}_\mathcal{R}}{\sqrt{2}} + \hat{p}_{\text{in}} + \hat{p}_\mathcal{S} - \frac{\hat{p}_{v\mathcal{R}}}{\sqrt{2}}, \label{eq:brau_clone_xp}
\end{eqnarray}
where the subscripts $C$, and $\text{``in''}$ denote the clone and the input state respectively.

\noindent \textbf{Reversible protocol: }
\begin{eqnarray}
\nonumber \hat{x}_{C} = \frac{\hat{x}_\mathcal{R} - \hat{x}_\mathcal{S}}{\sqrt{2}} + \hat{x}_{\text{in}} + \frac{\hat{x}_{v \mathcal{R}} - \hat{x}_{v \mathcal{S}}}{\sqrt{2}}, \\
\hat{p}_{C} = \frac{\hat{p}_\mathcal{R} + \hat{p}_\mathcal{S}}{\sqrt{2}} + \hat{p}_{\text{in}} - \frac{\hat{p}_{v \mathcal{R}} - \hat{p}_{v \mathcal{S}}}{\sqrt{2}}, \label{eq:peng_clone_xp} \\
\hat{x}_{aC} = \hat{x}_{\text{in}} - \sqrt{2}\hat{x}_{v \mathcal{S}}, ~~ \hat{p}_{aC} = -\hat{p}_{\text{in}} - \sqrt{2}\hat{p}_{v \mathcal{S}}. \label{eq:peng_anticlone_xp}
\end{eqnarray}
Here, $aC$ denotes the mode of the anticlone. Remarkably, in both protocols, the telecloning occurs symmetrically, i.e., all the clones are produced with the same fidelity. For Gaussian input states, resources, and unit gain, the fidelity is given readily in terms of the quadrature variances of the output state~\cite{Furusawa_Science_1998}:

\begin{eqnarray}
    \mathcal{F}=2/\sqrt{(1+ \langle \Delta^2 \hat{x}_{C} \rangle)(1+ \langle \Delta^2 \hat{p}_{C} \rangle)}. 
    \label{eq:fidelity}
\end{eqnarray}

\textcolor{black}{While we have restricted our analysis to the traditional telecloning protocol involving double-homodyne detection, minimal disturbance measurements in continuous variable setups have also been proposed \cite{Sabuncu_PRA_2007_2}. A protocol involving minimal disturbance measurements tends to reduce quantum noise and can thus be utilized to enhance the gain and fidelity in quantum communication protocols.}

\subsection{Variance-based non-classicality measure}
\label{sec:non-cl}

\begin{figure}
\centering
\includegraphics[width=0.8\linewidth]{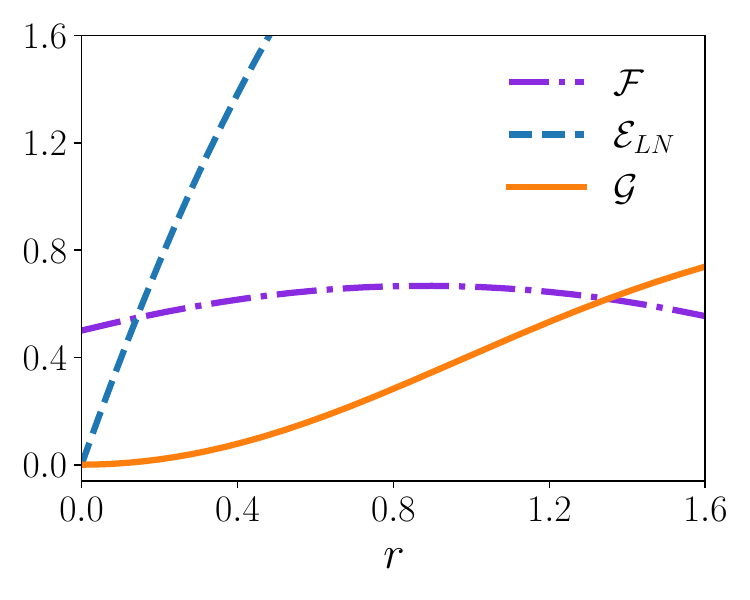}
\caption{\textbf{Fidelity and corresponding mode entanglement for coherent state telecloning with TMSV resource.} The fidelity $\mathcal{F}$ (violet dash-dotted line), the entanglement between the sender and the clone mode, $\mathcal{E}_{\text{LN}}$ (blue dashed line), and the genuine multimode entanglement $\mathcal{G}$ of the irreversible telecloning network (orange solid line) are shown (ordinate) against the squeezing strength $r$ of the resource (abscissa). Both the axes are dimensionless.}
\label{fig:moti} 
\end{figure}

When the TMSV state is used to generate two copies of the input at distant locations, the fidelity of the clones produced in the case of input coherent states is given by $\mathcal{F} = {4}/({5 + 3 \cosh 2 r -2 \sqrt{2} \sinh 2 r})$, where $r$ is the squeezing strength of the TMSV state. It is observable that the fidelity increases steadily from $\mathcal{F}_{\text{cl}} = \frac{1}{2}$ at $r = 0$ (which we refer to as the classical threshold, achievable in the absence of entanglement~\cite{Hammerer_PRL_2005}) up to $\mathcal{F}_{\max} = \frac{2}{3}$~\cite{Furusawa_Science_1998} at $r_{\text{opt}} \approx 0.881$. Importantly, the fidelity does not increase further upon increasing the squeezing strength, $r$.

In order to understand which specific quantum property of the resource drives the telecloning process, we study the bipartite entanglement as quantified by logarithmic negativity~\cite{Serafini_2017}, between the mode at the sender's station and the mode with one of the receivers. This reads as $\mathcal{E}_{\text{LN}} = - \log_2 \nu$, with $\nu = \frac{1}{16} \Big[3 + 4 \cosh 2 r + 9 \cosh 4 r - \sqrt{2(\sinh r - 3 \sinh  3 r)^2 (9 \cosh 2 r + 7)}\Big]$.~From the functional form, it is evident that $\mathcal{E}_{\text{LN}}$ increases monotonically with $r$, thereby failing to connect faithfully with the fidelity of telecloning, as shown in Fig.~\ref{fig:moti}. Furthermore, the genuine multimode entanglement, calculated using the generalized geometric measure (GGM)~\cite{Roy_PRA_2020}, between all three involved modes, cannot explain the trends of the fidelity either. Specifically, the GGM in this setup, $\mathcal{G} = 1 - \frac{2}{1 + \cosh^2 r}$, also increases with the squeezing amplitude, $r$, in contrast to $\mathcal{F}$ (see Fig.~\ref{fig:moti}). This leads us to conclude that neither bimodal entanglement nor genuine multimode entanglement can unambiguously explain the success of a telecloning protocol, despite entanglement being a necessary resource for the same. It can be argued that logarithmic negativity and GGM fail since they are designed in terms of the symplectic eigenvalues, which is not the case for fidelity. It also suggests that a different kind of non-classical property could be the resource for CV telecloning.

Let us concentrate on the separability criterion for two-mode CV states~\cite{Duan_PRL_2000}, which states that a bipartite state, $\rho_{ij}$, is necessarily separable if $\zeta_{i,j} = \langle \Delta^2 (\hat{x}_i - \hat{x}_j) \rangle + \langle \Delta^2 (\hat{p}_i + \hat{p}_j) \rangle \geq 1$, where $\langle \Delta^2(.) \rangle$ represents the variance. Considering the two modes to be those for the sender homodyne detection, $h$, and one of the clones, $C$, let us define a non-classicality measure for estimating the telecloning performance as

\begin{equation}
    \mathcal{Q}_{hC} = 1 - \zeta_{h, C}.
    \label{eq:non-cl-meas}
\end{equation}
Our conjecture for analyzing the performance of a telecloning protocol may now be stated as follows

\textbf{Proposition $\mathbf{1}$.} \textit{In a telecloning protocol involving Gaussian resource and input states, the fidelity of a clone can be estimated through the function $\mathcal{Q}_{hC} = 1 - \zeta_{h,C} = 1 - \langle \Delta^2 (\hat{x}_h - \hat{x}_C) \rangle - \langle \Delta^2 (\hat{p}_h + \hat{p}_C) \rangle$, where $(\hat{x}_h, \hat{p}_h)$ represent the position and momentum quadrature pair of the sender homodyne mode, $h$, and similarly for the clone mode, $C$. Non-classical fidelity is obtained when $0 \leq \mathcal{Q}_{hC} \leq 1$.} 

\textit{Proof}. We consider $C = \frac{\mathcal{R} + v\mathcal{R}}{\sqrt{2}}$ without loss of generality since both protocols are symmetric with respect to the clones produced. The homodyne mode $h = \mathcal{S}$ for the irreversible scheme and $h = \frac{\mathcal{S} + v\mathcal{S}}{\sqrt{2}}$ in the reversible case. Therefore, we obtain
\begin{eqnarray}
   \nonumber \langle \Delta^2(\hat{x}_h - \hat{x}_C) \rangle_{\text{irreversible}} = && \langle \Delta^2(\hat{x}_{\mathcal{S}}) \rangle + \frac{1}{2} \langle \Delta^2(\hat{x}_{\mathcal{R}}) \rangle \\ 
   && - \sqrt{2} \langle \hat{x}_{\mathcal{S}} \hat{x}_{\mathcal{R}} \rangle + \frac{1}{2},
\end{eqnarray}
\begin{eqnarray}
   \langle \Delta^2(\hat{x}_h - \hat{x}_C) \rangle_{\text{reversible}} = && \frac{1}{2} \Big( \langle \Delta^2(\hat{x}_{\mathcal{S}}) \rangle + \langle \Delta^2(\hat{x}_{\mathcal{R}}) \rangle \Big)\nonumber \\
   && - \langle \hat{x}_{\mathcal{S}} \hat{x}_{\mathcal{R}} \rangle + 1,
\end{eqnarray}
where we have used $\langle \Delta^2 \hat{x}_{v\mathcal{S}(\mathcal{R})} \rangle = 1$ for the vacuum state. Similar expressions can be obtained for $\langle \Delta^2(\hat{p}_h + \hat{p}_C) \rangle$ to construct $\zeta_{h,C}$. Using Eqs. \eqref{eq:brau_clone_xp} and \eqref{eq:peng_clone_xp}, we can readily obtain the clone quadrature variances and it follows that $\zeta_{h,C} = \langle \Delta^2 \hat{x}_C \rangle + \langle \Delta^2 \hat{p}_C \rangle - 2$ for both the protocols. The fidelity expression then becomes
\begin{eqnarray}
\mathcal{F} = \frac{2}{\sqrt{1 + \langle \Delta^2 \hat{x}_C \rangle + \langle \Delta^2 \hat{p}_C \rangle + \langle \Delta^2 \hat{x}_C \rangle \times \langle \Delta^2 \hat{p}_C \rangle}}.
\label{eq:fid_prop_1}
\end{eqnarray}
We can now consider that $\zeta_{h,C} = \zeta_{h,C}^x + \zeta_{h,C}^p$, with $\zeta_{h,C}^q = \langle \Delta^2 \hat{q}_C \rangle - 1$ with $q  = \{x,p\}$. Therefore, the last term in the denominator of Eq. \eqref{eq:fid_prop_1} becomes $\langle \Delta^2 \hat{x}_C \rangle \times \langle \Delta^2 \hat{p}_C \rangle = 1 + \zeta_{h,C}^x + \zeta_{h,C}^p + \zeta_{h,C}^{xp}$, where we have denoted $\zeta_{h,C}^x \times \zeta_{h,C}^p = \zeta_{h,C}^{xp}$. Putting everything together in terms of $\zeta_{h,C}$ and translating Eq. \eqref{eq:fid_prop_1} in terms of $\mathcal{Q}_{hC}$, we can approximate the clone fidelity as
\begin{eqnarray}
    \mathcal{F} = \frac{2}{\sqrt{6 - 2 \mathcal{Q}_{hC} + \zeta_{h,C}^{xp}}}.
\end{eqnarray}
For $\mathcal{Q}_{hC} \in [0,1]$, we must have $0 \leq \zeta_{h, C}^{x} + \zeta_{h,C}^{p} \leq 1 \implies 0 \leq \zeta_{h,C}^{xp} \leq 1$. Therefore, 
\begin{eqnarray}
    \frac{2}{\sqrt{7 - 2 \mathcal{Q}_{hC}}} \leq \mathcal{F} \leq \frac{2}{\sqrt{6 - 2 \mathcal{Q}_{hC}}},
\end{eqnarray}
and we obtain $\mathcal{F} \geq 1/2$ when $\mathcal{Q}_{hC}$ is positive semi-definite. Hence the proof. $\hfill \blacksquare$

\textbf{Remark.} As we will demonstrate, such a connection also remains true for non-Gaussian resource states. Note that a closed form of the fidelity expression, as in Eq. \eqref{eq:fidelity}, is not available for the non-Gaussian case, although it can be taken as an approximation in the limit of small non-Gaussianity or when the higher order quadrature moments are negligible as compared to the corresponding variance.

\begin{figure}[ht]
\includegraphics[width=0.8\linewidth]{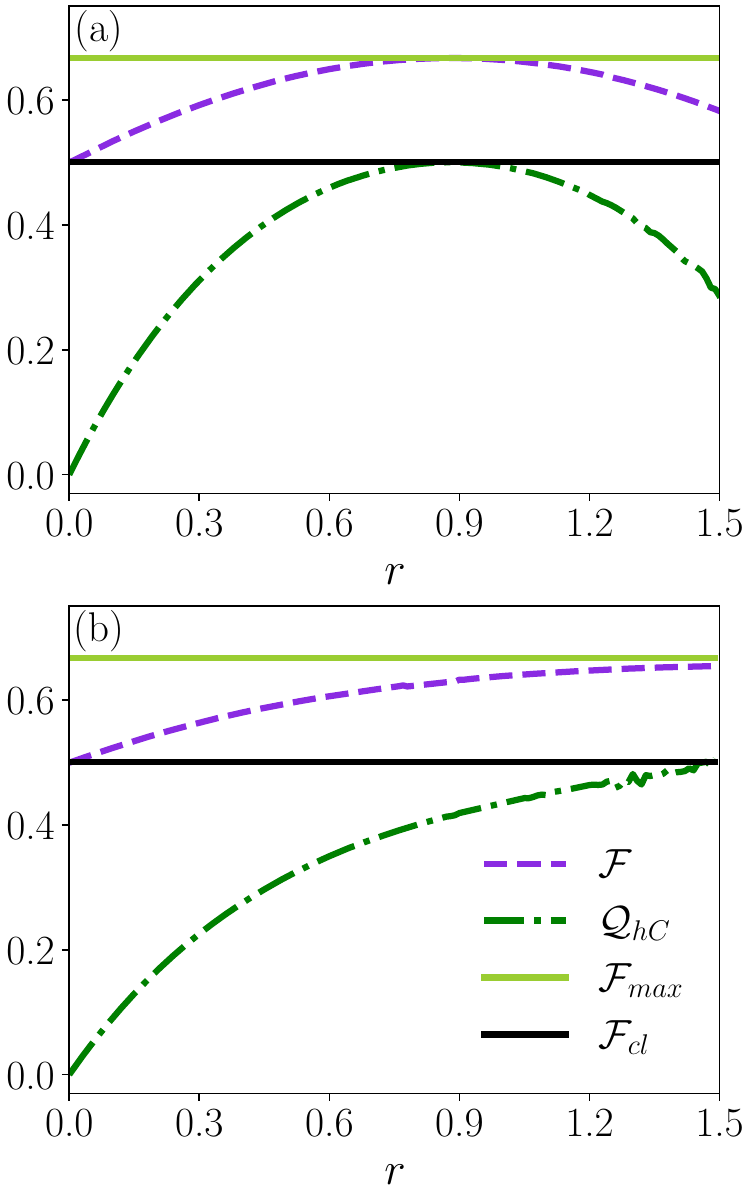}
\caption{\textbf{Fidelity and non-classicality measure for telecloning of coherent states using TMSV resource.} The clone fidelity $\mathcal{F}$ (violet dashed line) and the non-classicality parameter $\mathcal{Q}_{hC}$ (dark green dash-dotted line), of the modes $h$ (sender's homodyne) and $C$ (clone mode at receiver station), are shown (ordinate) with respect to the resource squeezing $r$ (abscissa) for (a) irreversible and (b) reversible protocols. The classical benchmark $\mathcal{F}_{\text{cl}} = 0.5$ (black horizontal solid line) and $\mathcal{F}_{\max} = 2/3$ (light green horizontal solid line) are also plotted for reference. Both axes are dimensionless.}
\label{fig:qex} 
\end{figure}

Since $\zeta_{h,C} < 1$ implies that the two modes may be inseparable, the above measure links the protocol's performance to the entanglement between the sender's and the clone's modes, albeit in a qualitative manner. We justify our claim with the help of Figs.~\ref{fig:qex}(a) and (b), where we consider both the reversible ($\mathcal{F} = \frac{2}{3 + e^{-2r}}$~\cite{Zhang_PRA_2006})  and irreversible ($\mathcal{F} = \frac{4}{5 + 3 \cosh 2 r -2 \sqrt{2} \sinh 2 r}$~\cite{van-Loock_PRL_2001}) telecloning schemes. We observe that as long as $\mathcal{Q}_{hC} > 0$, the clone is produced with non-classical fidelity, i.e., $\mathcal{F} > \frac{1}{2}$ in the irreversible and reversible schemes. 
We notice that, even though the two protocols differ in their approach to preparing the clones, the respective fidelities can be gauged with the help of $\mathcal{Q}_{hC}$ as shown in Fig.~\ref{fig:qex}. Furthermore, a higher value of $\mathcal{Q}_{hC}$ indicates a higher fidelity of the clone produced in the mode $C$. We note also that since $\zeta$ is a necessary condition for separability, $\mathcal{Q} < 0$ might also indicate the presence of entanglement. However, in that scenario, $\mathcal{F}$ is always classical, thereby illustrating that entanglement, although necessary, is not sufficient to guarantee quantum advantage in the telecloning protocol.

Note that the measure $\mathcal{Q}$ is not merely a mathematical construct.
In particular, the quantity $\zeta$ is closely related to an operational measure of non-classicality~\cite{Gehrke_PRA_2012}, specifically in terms of quadrature squeezing. Recall that squeezing reduces the quadrature noise below the vacuum level, which is an innate quantum property. Moreover, $\zeta$ has also been used to construct a witness in the operational resource theory of continuous variable nonclassicality~\cite{Yadin_PRX_2018}, and is experimentally discernable using quantum noise-free measurements~\cite{Gehrke_PRA_2012}. It is also shown to be related with monogamy of entanglement \cite{Rosales_PRA_2017} in CV systems, based on measurements of the quadrature variables. 
Our measure, $\mathcal{Q}$ indicates that nonclassicality serves as an essential resource in continuous variable telecloning protocols (cf. \cite{Das_PRA_2023} for discrete variable systems). In the succeeding sections, we use $\mathcal{Q}$ to investigate telecloning using non-Gaussian resource states.

\textbf{Remark:} It is evident that in the irreversible protocol, the Gaussian TMSV state attains maximum clone fidelity for $r < 1.0$. In the reversible protocol, on the other hand, the fidelity with TMSV resource increases monotonically with $r$ and attains optimality only in the limit of infinite squeezing. Given these observations, we shall restrict our analysis of telecloning up to $r = 1.0$, so that non-Gaussian advantage can be properly studied for small squeezing strengths wherein Gaussian resource states provide sufficiently high clone fidelities.

\section{Non-Gaussian advantage in CV telecloning}
\label{sec:n-G_tele}

Although telecloning of continuous-variable states, specifically coherent states, has been demonstrated using the Gaussian TMSV state as well as $SU(m,1)$ coherent states~\cite{Ferraro_PRA_2005} and Gaussian valence bond states~\cite{Adesso_OS_2007}, the performance of non-Gaussian states in the protocol has not been explored. To this end, we employ photon-addition and -subtraction in one or both the modes of the TMSV state, to generate non-Gaussian resources. \textcolor{black}{Creation of non-Gaussian states using photon-addition and subtraction have already been experimentally demonstrated~\cite{Dakna_EPJD_1998, Parigi_Science_2007, Ourjoumtsev_PRL_2007, Takahashi_NP_2010, Kurochkin_PRL_2014, Loaiza_NPJ_2019, Zhang_PRA_2006, Thapliyal_PRR_2024}.} For non-Gaussian states, we employ the Wigner function formalism to evaluate the success of the telecloning protocol. Recall that Born's rule provides the framework for the achievable fidelity of the cloned state, $\rho_C$, with Wigner function $W_C(x_C, p_C)$, against the input state, $\rho_\text{in}$, having corresponding Wigner function $W_\text{in}(x_\text{in}, p_\text{in})$, as~\cite{Adesso_OSID_2014, Barnett_2002}
\textcolor{black}{
\begin{eqnarray}
\nonumber \mathcal{F}=\Tr [\rho_{\text{in}} \rho_C] = && 2\pi  \int dx_C dp_C dx_{\text{in}} dp_{\text{in}} W_C(x_C, p_C) \times \\
&& \nonumber W_\text{in}(x_\text{in}, p_\text{in}) \delta(x_C - x_{\text{in}}) \delta(p_C - p_{\text{in}}) \\
  = 2\pi \int dx_{\text{in}} dp_{\text{in}} && W_C (x_{\text{in}}, p_{\text{in}}) W_\text{in}(x_\text{in}, p_\text{in}), \label{eq:fid_int}
\end{eqnarray}
}
where $\delta(.)$ denotes the delta function. See Appendix~\ref{app:cv} for a primer on CV phase space formalism. \textcolor{black}{The fidelity of the cloned states using non-Gaussian resources is calculated by numerically integrating Eq. \eqref{eq:fid_int}.}

\subsection{Telecloning of coherent states using non-Gaussian resources}
\label{subsec:tele_coh_nG}

\begin{figure*}[ht]
\includegraphics[width=0.7\linewidth]{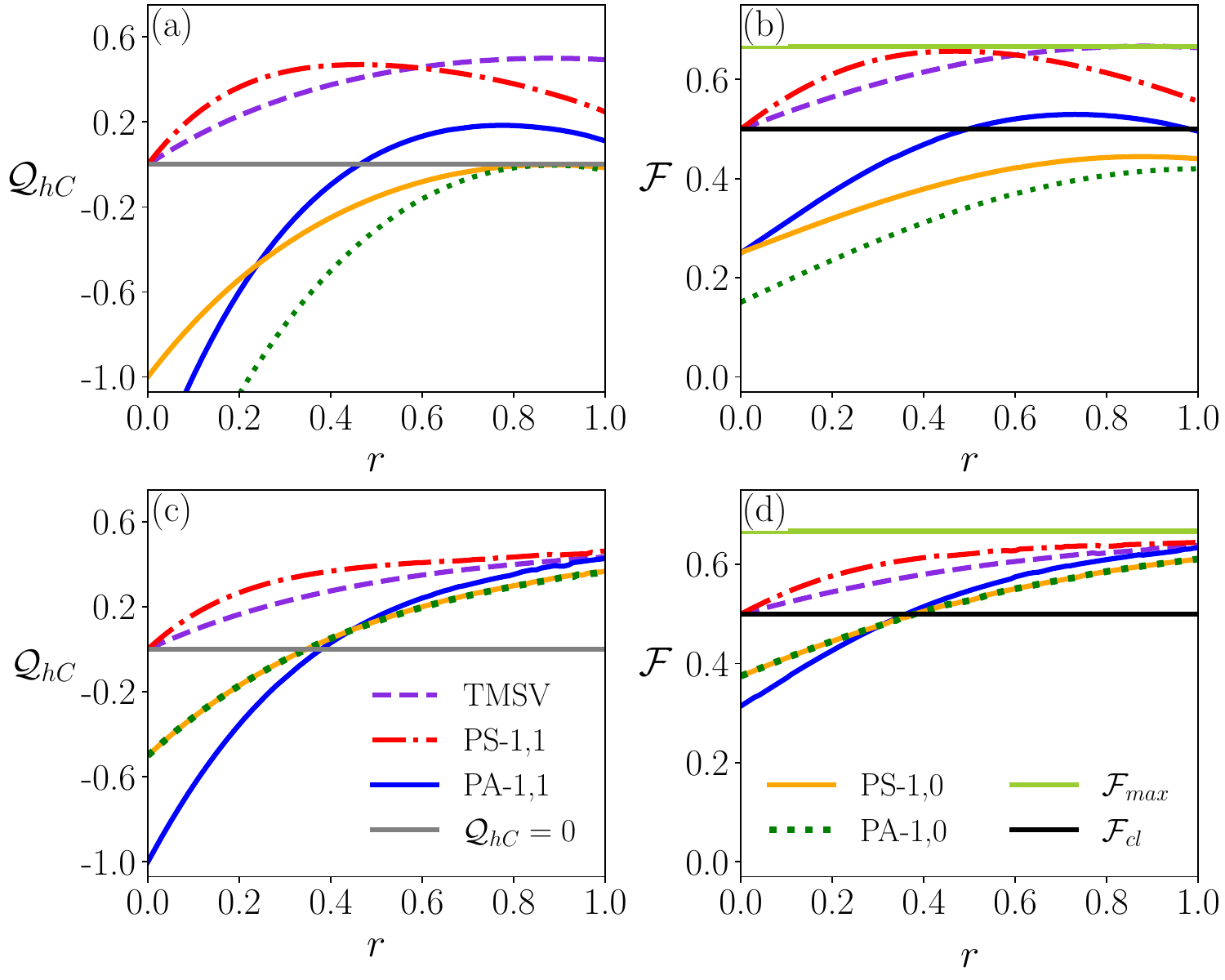}
\caption{\textbf{Fidelity and nonclassicality measure with Gaussian and non-Gaussian resources.} The coherent state is taken as the input for telecloning. (Upper panel) \textit{Irreversible protocol}. (a) The nonclassicality measure $\mathcal{Q}_{h C}$ (ordinate), between the sender homodyne mode $h$ and the clone mode $C$, versus the resource squeezing $r$ (abscissa) for TMSV (violet dashed line), PS-$1,1$ (red dash-dotted line), PA-$1,1$ (dark blue solid line), PS-$1,0$ (orange solid line), and PA-$1,0$ (green dotted line) resource states. $\mathcal{Q} = 0$ (grey horizontal solid line) is shown for reference. (b) The fidelity, $\mathcal{F}$ (ordinate), of a clone is illustrated against the squeezing strength $r$ (abscissa) for the same resource states as in (a). The classical fidelity $\mathcal{F}_{\text{cl}} = 0.5$ (black horizontal solid line) and the maximum fidelity $\mathcal{F}_{\max} = 2/3$ (green horizontal solid line) are shown for reference. (Lower panel) \textit{Reversible protocol}. (c) $\mathcal{Q}_{h C}$ (ordinate) against $r$ (abscissa) and (d) $\mathcal{F}$ (ordinate) against $r$ (abscissa). All specifications are the same as in the upper panel. Both axes are dimensionless. }
\label{fig:coh} 
\end{figure*}

In order to demonstrate the advantage of incorporating non-Gaussianity in the telecloning scheme, four different non-Gaussian states have been taken into account, viz., the two-mode photon-added state (PA-$n,n$), the two-mode photon-subtracted state (PS-$n,n$) and the single-mode photon-added (PA-$n,0$) and photon-subtracted (PS-$n,0$) states. Here, $n$ represents the number of photons added (subtracted) to (from) one or both modes. Note that the state with $n$ photons added in a single mode is operationally equivalent to the one with $n$ photons subtracted from the other mode~\cite{Roy_PRA_2018}. \textcolor{black}{It must be mentioned here that from an experimental point of view, generation of such resources is an inherently probabilistic task  \cite{Takahashi_NP_2010, Navarrete_PRA_2012, Kurochkin_PRL_2014, Loaiza_NPJ_2019, Kumar_PRA_2022} as the process of adding (subtracting) photons to (from) each mode of the TMSV state is not deterministic.} Following the Wigner functions representation of these paradigmatic non-Gaussian states~\cite{Kumar_PRA_2022}, it is straightforward to derive the Wigner function $W_C(x_C,p_C)$ for the clones state and subsequently its fidelity, once the same for the input and resource states are known.  In Appendix~\ref{app:tele_wig}, we provide an outline of the calculations involved in obtaining the telecloning fidelity using Wigner functions.

During the telecloning of coherent states using the irreversible protocol, clones with optimal fidelity are obtained when the excess noise in the clone quadratures, $\hat{x}_C$ and $\hat{p}_C$, are symmetric~\cite{van-Loock_PRL_2001}. This is possible when a vacuum state impinges on the balanced beam splitter, which is used to split the modes among the distant receivers.
For the irreversible protocol, we can observe that the PS-$1,1$ non-Gaussian state furnishes the highest maximum fidelity among all the considered non-Gaussian resources, $\mathcal{F}_{\max}^{\text{PS-}1,1} \approx 0.656$ at $r_{\text{opt}}^{\text{PS-}1,1} \approx 0.4137$ (here, we define $r_{\text{opt}}^{\rho}$ as the squeezing amplitude at which the resource state $\rho$ provides the maximum fidelity). This is, however, lower than the maximum fidelity $\mathcal{F}_{\max}^{\text{TMSV}} = 0.667$, obtained through the TMSV resource at $r_{\text{opt}}^{\text{TMSV}} = 0.881$ as shown in Fig.~\ref{fig:coh}(b).

The advantage of non-Gaussianity is apparent on two fronts - $(i)$ below $r \approx 0.5$, the photon-subtracted state always dominates over the Gaussian TMSV state in providing clones with higher fidelity, and $(ii)$ the maximum fidelity for the non-Gaussian state is attained at much lower squeezing amplitude, $r$. Conversely, the PA-$1,1$ non-Gaussian state provides no advantage compared to the TMSV state, being able to implement quantum advantage $(\mathcal{F} \geq 0.5)$ only in a limited range of $r \in (0.5,1)$. The non-Gaussian states obtained by adding (subtracting) a single photon to (from) a single mode of the TMSV state perform the worst since they cannot even overcome the classical threshold, over the entire range of squeezing. 
The situation is qualitatively similar in the reversible protocol (see Fig.~\ref{fig:coh}(d)). The PS-$1,1$ state provides a clear advantage over its Gaussian counterpart, although, the maximum fidelity of $2/3$ is obtained only at infinite squeezing. No other non-Gaussian state, like the PA-$1,1$, PA-$1,0$, or PS-$1,0$, can outperform the TMSV resource, even though some of them can provide quantum advantage by beating the classical threshold for moderate resource squeezing $r \gtrsim 0.4$. Finally, we note that Eq. \eqref{eq:peng_anticlone_xp} indicates that the quadratures corresponding to the anticlone are independent of the resource state. Therefore, both Gaussian and non-Gaussian states produce anticlones with the same classical fidelity of $1/2$~\cite{Zhang_PRA_2006}. 
\textcolor{black}{It is important to note that we have considered $\mathcal{F} = 2/3$ as the quantum benchmark for coherent state telecloning. However,  there can exist nonuniversal, local cloning maps for higher-dimensional or CV states that can achieve fidelity greater than universal cloning machines~\cite{Rafal2004, Cerf_PRL_2005}. Unlike local cloning maps, no amplification of the input state is needed and the protocols considered here rely only on squeezed states and linear optics~\cite{van-Loock_PRL_2001}. So far, using several Gaussian and non-Gaussian resources, such as two-mode squeezed vacuum (TMSV), photon-added and subtracted TMSV states (including single and multi-photon operations), and other bimodal superposition states, we have not observed fidelities above 2/3.} Let us now investigate how nonclassicality, as estimated by our proposed measure $Q$, performs as the key resource for the telecloning of coherent states.

\subsection{Quadrature variance as an indicator of coherent state telecloning fidelity}
\label{subsubsec:coh_nc}

\begin{figure*}[ht]
\includegraphics[width=0.99\linewidth]{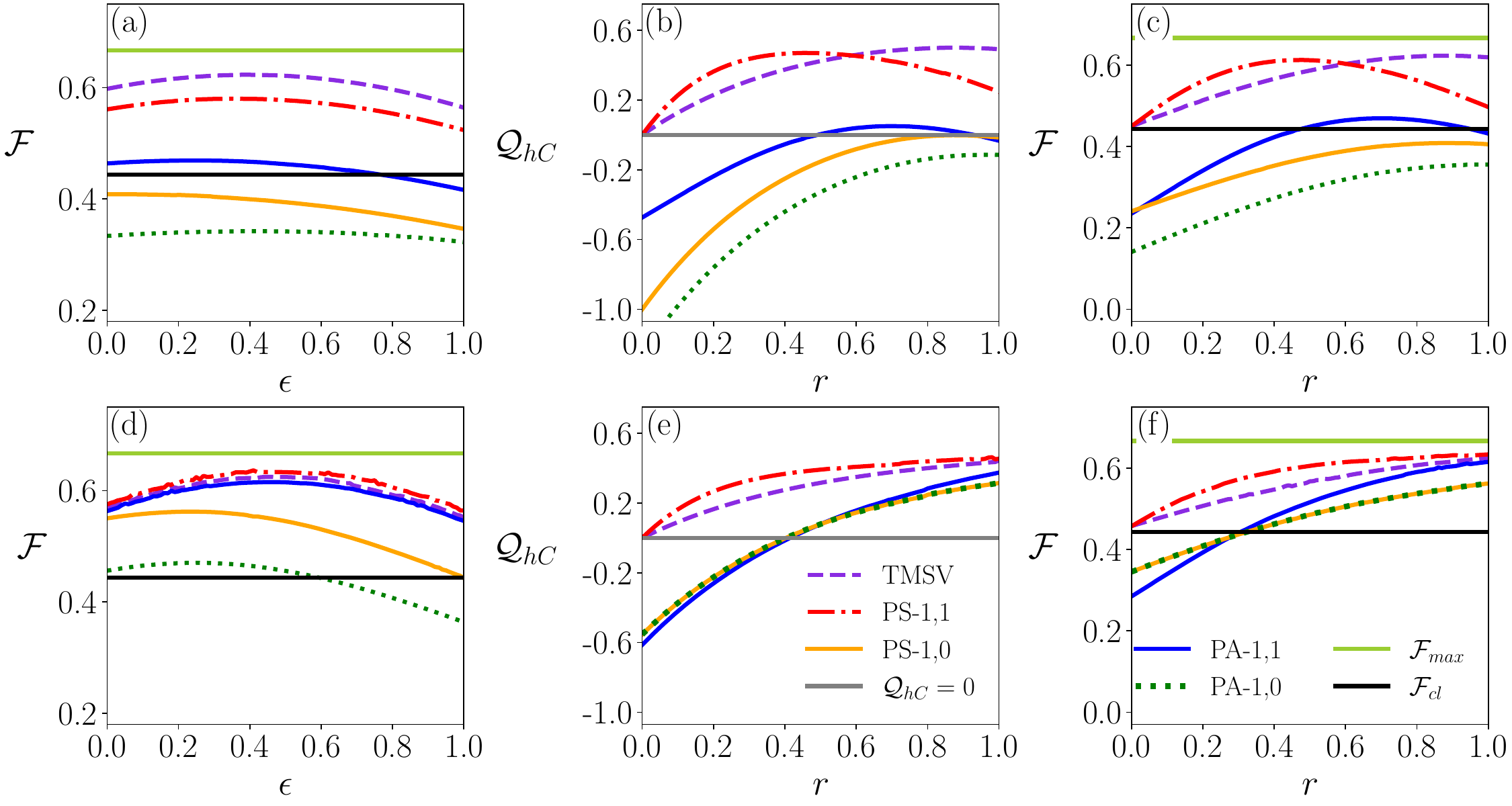}
\caption{\textbf{Fidelity and nonclassicality measure for telecloning of squeezed input states with Gaussian and non-Gaussian resources.} (Upper panel) \textit{Irreversible protocol} (a) The fidelity $\mathcal{F}$ (y-axis) for telecloning a squeezed state of squeezing parameter $s = 0.5$ is shown versus the squeezing $\epsilon$ (x-axis) of the state impinged on the balanced beam splitter. (b) The nonclassicality measure $\mathcal{Q}_{h C}$ (y-axis) is plotted against the resource squeezing $r$ (x-axis). (c) The fidelity $\mathcal{F}$ (y-axis) is shown against the resource squeezing $r$ (x-axis). (Lower panel) \textit{Reversible protocol} (d), (e), and (f) are the same as (a), (b), and (c). All specifications are the same as in Fig.~\ref{fig:coh}, except that $\mathcal{F}_{\text{cl}} \approx 0.44$. Both axes are dimensionless. }
\label{fig:sq} 
\end{figure*}

For both the irreversible and the reversible protocols, we need to consider the nonclassicality measure concerning the sender homodyne mode $h$ and the mode holding one of the clones, say, mode $C$. Since the protocol produces symmetric clones, $\mathcal{Q}_{h C}$ is enough to gauge the fidelity of all the produced clones. 
In the irreversible protocol, $\mathcal{Q}_{h C}$ assumes its maximum value for the TMSV state at $r \approx 0.881$, which is exactly the squeezing amplitude where $\mathcal{F}_{\max}^{\text{TMSV}} = 2/3$. On the other hand, the maximum of $\mathcal{Q}_{hC}$, which we denote as $Q_{h C}^{\max}$, for the PS-$1,1$ state occurs at a much lower value of $r$, which also resembles the behavior of the fidelity. However, as illustrated in Fig.~\ref{fig:coh}(a), $Q_{{h C}}^{\max}(\text{PS-}1,1) < Q_{{h C}}^{\max}(\text{TMSV})$, which explains why the photon-subtracted state cannot generate a higher optimal fidelity as compared to the Gaussian resource. The PA-$1,1$ state exhibits nonclassicality within $0.5 \lesssim r \lesssim 1.0$, but can never overcome $\mathcal{Q}_{h C}(\text{TMSV})$, and hence cannot provide a definite non-Gaussian advantage in the telecloning protocol. The single photon-added (subtracted) non-Gaussian states only manifest classical properties, as quantified by $\mathcal{Q}_{h C}$, and are, therefore, deemed useless for the protocol. 

Considering the reversible protocol, we observe that both the TMSV state and the PS-$1,1$ state are strictly nonclassical for any non-vanishing squeezing parameter as depicted in Fig.~\ref{fig:coh} (c). Furthermore, unlike the irreversible scheme, $\mathcal{Q}_{{h C}}(\text{PS-}1,1) > \mathcal{Q}_{{h C}}(\text{TMSV})$, for the entire range of $r$. Expectedly, we find that $\mathcal{F}^{\text{TMSV}} < \mathcal{F}^{\text{PS}-1,1}$ (see Fig.~\ref{fig:coh} (d)). In this situation, all the non-Gaussian states, PA-$1,1$ and PA(S)-$1,0$, can furnish quantum advantage for $r \gtrsim 0.4$ although none can perform better than the TMSV state. Again, the squeezing amplitude, $r$, above which $\mathcal{F} > \mathcal{F}_{\text{cl}}$ occurs matches with the one that can be obtained from positive $\mathcal{Q}_{hC}$. Notice that, in this case, $\mathcal{Q}_{hC}$ can provide a threshold on $r$ beyond which $\mathcal{Q}_{hC} > 0 \implies \mathcal{F} > \mathcal{F}_{\text{cl}}$ since the fidelity is monotonic with $r$. Such a threshold on $r$ cannot be provided for the irreversible protocol due to the non-monotonic nature of the fidelity function, although an estimate on $r$ can be faithfully recovered through the behavior of $\mathcal{Q}_{hC}$. Therefore, the connection established for Gaussian input and resource states between $\mathcal{Q}_{hC}$ and $\mathcal{F}$ remains valid even for non-Gaussian resource states with coherent state inputs in both protocols.

\textcolor{black}{\textbf{Remark.} The telecloning protocol also demands proper processing of the input coherent states. The techniques demonstrated in Ref. \cite{Sabuncu_PRA_2010} enable the protection of the coherent state during telecloning, ensuring improved preservation of input information, which is of paramount application in quantum key distribution (QKD). Additionally, environmental measurements can assist
in correcting quantum information~\cite{Filip_PRA_2005, Sciarrino_PRA_2009} when using non-Gaussian resources, which can improve error tolerance. This is of great relevance in telecloning where one lacks complete control over environmental conditions.}

\subsection{Non-Gaussian telecloning of input squeezed states}
\label{subsec:sq_ng}

\begin{figure}[ht]
\includegraphics[width=0.9\linewidth]{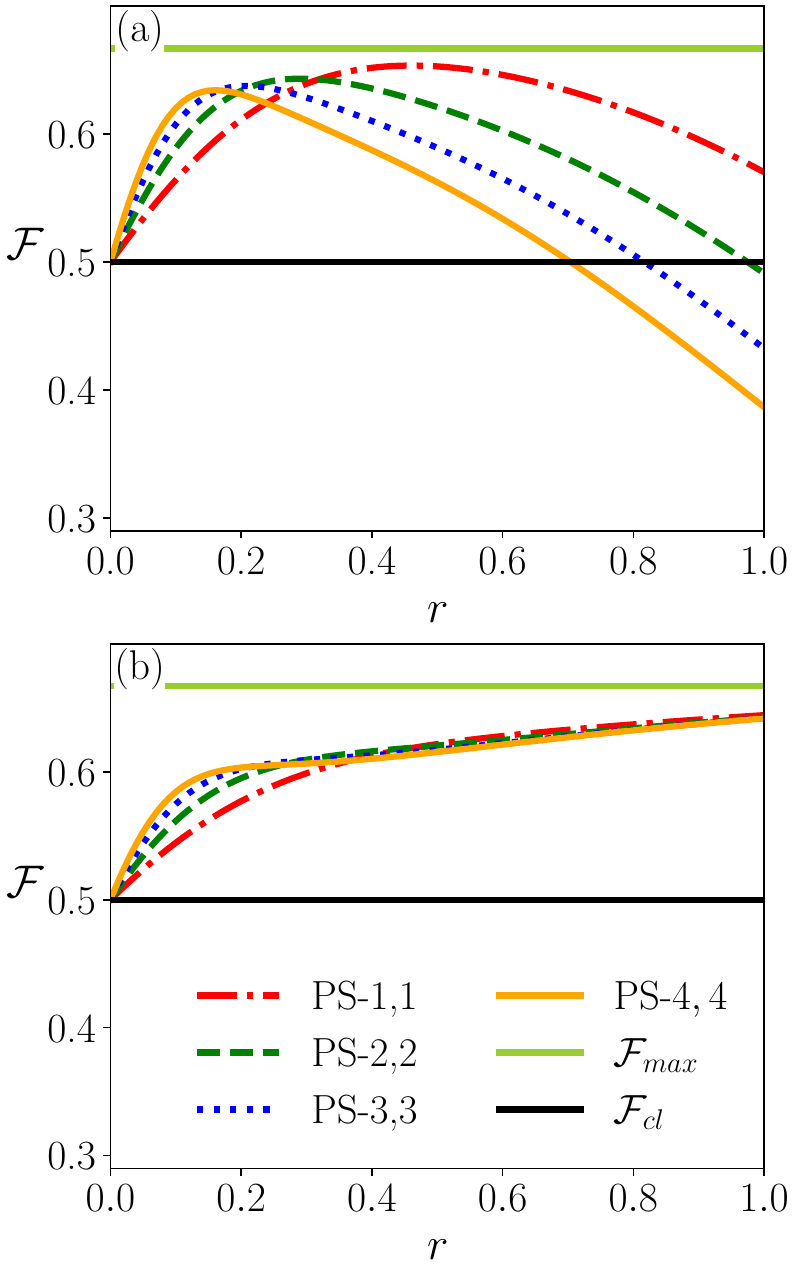}
\caption{\textbf{The fidelity of telecloning input coherent states with higher-order non-Gaussian photon-subtracted states.} The fidelity $\mathcal{F}$ (ordinate) is illustrated against the resource squeezing $r$ (abscissa) for PS-$1,1$ (red dash-dotted line), PS-$2,2$ (green dashed line), PS-$3,3$ (blue dotted line), and PS-$4,4$ (orange solid line) resource states in (a) irreversible and (b) reversible protocols. $\mathcal{F}_{\text{cl}} = 0.5$ (grey horizontal solid line) and $\mathcal{F}_{\max} = 2/3$ (green horizontal solid line) are also shown for reference. Both axes are dimensionless. }
\label{fig:ps} 
\end{figure}

Contrary to the telecloning of coherent states, optimal fidelity of squeezed state clones is obtained through asymmetric excess noise in the quadratures of the clone modes~\cite{van-Loock_PRL_2001}. This constraint demands that, when preparing the multimode network, we split the receiver modes through squeezed state inputs at the balanced beam splitter. With the vacuum state impinged on the beam splitter, the nonclassicality measure is the same as discussed in the previous section (see Figs.~\ref{fig:coh} (a) and (c)). To estimate the impact of generating the telecloning network using squeezed states, we study the variation of $\mathcal{F}$ with respect to $\epsilon$, which we define to be the squeezing strength of the input at the beam splitter at the receiver station (see Fig.~\ref{fig:sch} where it is denoted as $\hat{S}(\epsilon)$). Note that \(\epsilon =0\) represents the vacuum state. Through this analysis, we try to estimate the optimal squeezing amplitude, $\epsilon_{\text{opt}}$, of the states impinged on the beam splitter, for which the fidelity of the clones would be near maximum for squeezed input states. To do so, we fix the squeezing parameter, $r$, of the various resource states at the value where they attain maximum $\mathcal{Q}_{h C}$ for coherent state telecloning: $r_{\text{TMSV}} = 0.89, r_{\text{PS}-1,1} = 0.47, r_{\text{PA}-1,1} = 0.73, r_{\text{PS}-1,0} = 0.88, r_{\text{PA}-1,0} = 0.02$ for the irreversible protocol and $r = 1.0$ for the reversible protocol. Note that, we consider the input state squeezing amplitude as $s = 0.5$.  The constructive effect of squeezed input to the beam splitter is tangible from Figs.~\ref{fig:sq} (a) and (d) wherein $\mathcal{F}$ increases with $\epsilon$, becoming maximum at $\epsilon  = \epsilon_{\text{opt}}$ for the resource states in both the protocols. Therefore, for each Gaussian or non-Gaussian state, the fidelity of the clones will be higher at $\epsilon_{\text{opt}}$, as compared to $\epsilon = 0$. Table.~\ref{tab:e_max} enumerates $\epsilon_{\text{opt}}$ for the different resource states for both the considered protocols. 
\textcolor{black}{Note that the generation of squeezed states is experimentally challenging. There, however, exist averaging protocols \cite{Lassen_PRA_2010} which allow for stabilizing the squeezing degree of squeezed state resources and are thus fundamental for our task. Furthermore, the distillation of squeezed states using photon-subtraction mechanism has also been demonstrated \cite{Dirmeier_OE_20202020, Grebien_PRL_2022}.}

\begin{table}[] 
	\begin{tabular}{|r|rr|}
\hline
\multicolumn{1}{|l|}{} & \multicolumn{2}{c|}{$\epsilon_{\text{opt}}$}                \\ \hline
\multicolumn{1}{|l|}{} & \multicolumn{1}{c|}{irreversible} & \multicolumn{1}{c|}{reversible} \\ \hline
TMSV                   & \multicolumn{1}{r|}{0.38}       & 0.45                      \\ \hline
PS-$1,1$                & \multicolumn{1}{r|}{0.33}       & 0.46                      \\ \hline
PA-$1,1$                & \multicolumn{1}{r|}{0.22}       & 0.46                      \\ \hline
PS-$1,0$                & \multicolumn{1}{r|}{0.02}       & 0.25                      \\ \hline
PA-$1,0$                & \multicolumn{1}{r|}{0.42}       & 0.23                      \\ \hline
\end{tabular}
\caption{$\epsilon_{\text{opt}}$ for telecloning squeezed states with squeezing $s = 0.5$ using the irreversible and the reversible protocols.\\}
\label{tab:e_max}
\end{table}

Comparing Figs.~\ref{fig:sq} (b) and (e) with Figs.~\ref{fig:sq} (c) and (f) respectively, we observe that the relation between $\mathcal{F}$ and $\mathcal{Q}_{hC}$ remains the same as in the previous situation with coherent state inputs. Specifically, $\mathcal{F}$ and $\mathcal{Q}_{hC}$ are non-monotonic with $r$ for the irreversible case, while a monotonic nature appears for the reversible one. Note that, unlike the coherent state scenario, $\mathcal{F}_{\text{cl}} \approx 0.442$ in this case, which is overcome by the PS-$1,1$, PA-$1,1$ and the TMSV states in the irreversible protocol, while all the considered Gaussian and non-Gaussian states considered here provide quantum advantage in the reversible scheme. However, only the photon-subtracted state, PS-$1,1$, can outperform the TMSV state.

\textcolor{black}{ Although we have not considered any imperfections in the telecloning setup, practical realizations can lead to the reduction of the efficacy of the nonclassical resource and the telecloning protocol. Taking into account the possible detector inefficiencies, it was demonstrated that one can suitably modulate the gain factors in the telecloning protocol to obtain optimum clones~\cite{Sabuncu_PRA_2008}.}


\subsection{Telecloning with higher-order non-Gaussian states}

Let us briefly inspect the benefits rendered by higher-order non-Gaussian states, especially states generated through subtraction of an equal number of photons from both modes of an initial TMSV state, i.e., PS-$n,n$. Our choice stems from the fact that in both the reversible and irreversible telecloning protocols, we have already shown that PS-$1,1$ is the best among all non-Gaussian states. For input coherent states and the irreversible scheme, higher fidelity at smaller resource squeezing is provided by the states with a larger number of subtracted photons, as seen in Fig.~\ref{fig:ps}. While this may prove to be an advantage when sufficient squeezing is inaccessible, it must be noted that with increasing $n$, the maximum fidelity value decreases. Therefore, higher-order non-Gaussian PS states cannot provide any constructive advantage in the irreversible telecloning protocol. In the case of the reversible scheme, although PS-$n,n$ states can always offer an advantage over the TMSV state, their fidelity again decreases with $n$ for high resource squeezing. Further, the maximum fidelity, $\mathcal{F}_{\max} = 2/3$ is still achieved only when $r \to \infty$. Therefore, we can conclude that creating higher-order non-Gaussian PS states is only favorable for the telecloning network in the limit of low squeezing strength in the shared initial state. It should be noted that the probability of successfully creating such states in experiments decreases drastically with an increase in the number of subtracted photons (see Table. $1$ in Ref.~\cite{Loaiza_NPJ_2019}).

\section{Asymmetric telecloning network}
\label{sec:asymm_tele}

\begin{figure*}
\includegraphics[width=0.7\linewidth]{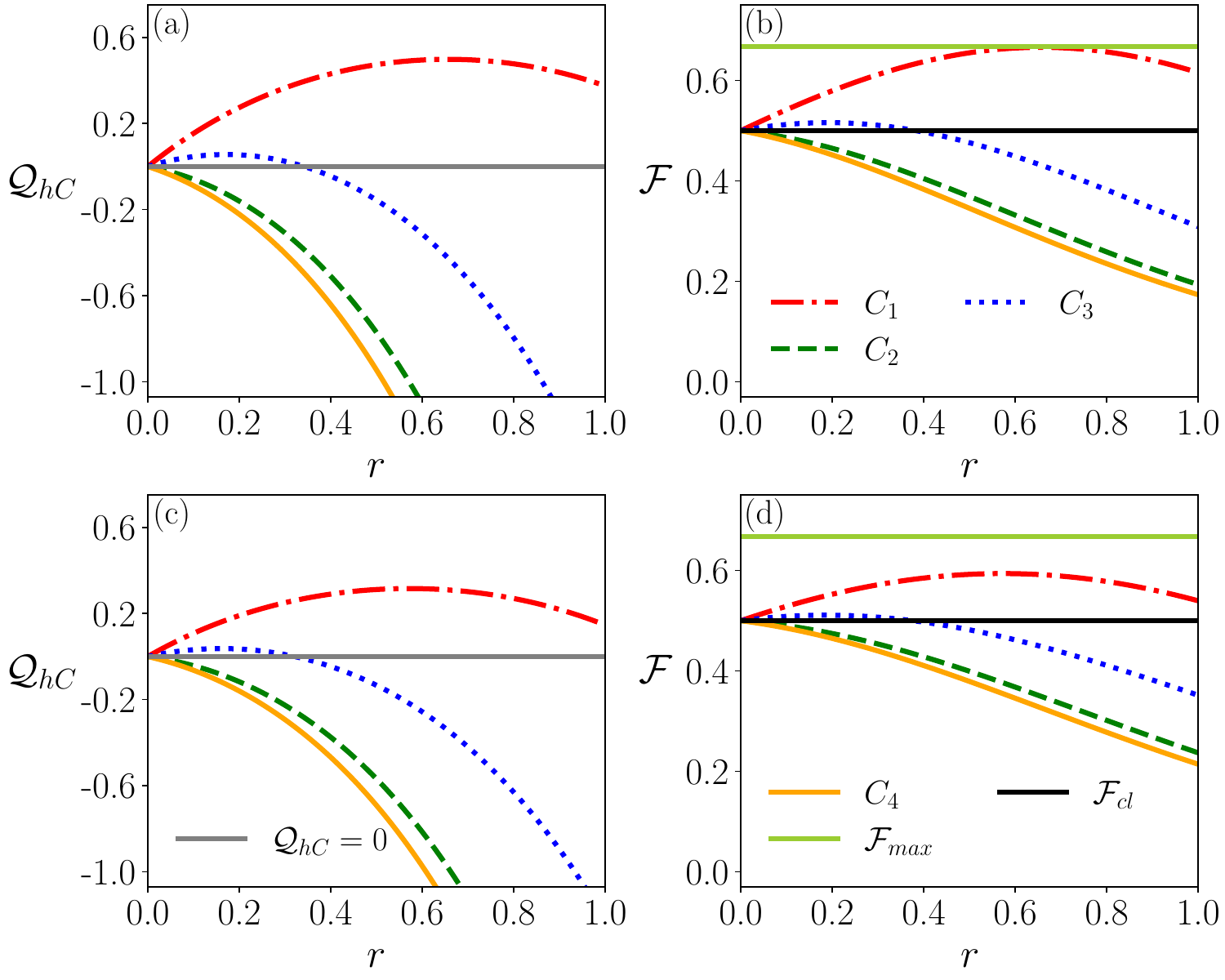}
\caption{\textbf{The nonclassicality measure and the corresponding clone fidelities for asymmetric telecloning of input coherent states.} (Upper panel) \textit{Irreversible protocol} $\mathcal{Q}_{hC_i}$ (a) and the fidelity $\mathcal{F}$ (b) (ordinate) against the resource squeezing $r$ (abscissa) for clones $C_1$ (red dash-dotted line), $C_2$ (green dashed), $C_3$ (blue dotted), and $C_4$ (orange solid line) using a five-mode genuinely entangled Gaussian resource state. (Lower panel) \textit{Reversible protocol} $\mathcal{Q}_{hC_i}$ (c) and the fidelity $\mathcal{F}$ (d) are shown with the same specifications as in (a) and (b) respectively. The classical benchmark $\mathcal{F}_{\text{cl}} = 0.5$ (dark black horizontal solid line), the maximum allowed fidelity $\mathcal{F}_{\max} = 2/3$ (light green horizontal solid line), and $\mathcal{Q}_{hC_i} = 0$ (light grey solid line) are plotted for reference. Both axes are dimensionless. }
\label{fig:asymm} 
\end{figure*}

When a two-mode Gaussian or non-Gaussian state is used to create the telecloning network, all the clones produced at the receiver modes have the same fidelity in the case of both the irreversible and the reversible protocols. Therefore, the entire discussion so far concerns symmetric telecloning schemes. But what if Alice wants to prioritize some receivers over others, for example, in case one or more receivers are untrustworthy?  To deal with such a situation, one must consider the asymmetric telecloning protocol~\cite{Murao_PRA_2000, Zhang_PRA_2006}. Note that, in the discrete variable (qudit) regime, such analysis has been widely performed~\cite{Cerf_PRL_2000, Cerf_JMO_2000, Murao_PRA_2000, Ghiu_PRA_2003, Zhao_PRL_2005}. Here, we propose an asymmetric telecloning setup for the continuous variable paradigm. One efficient way to do this is to create an asymmetric $N+1$-mode genuinely entangled resource state and distribute it among the sender and $N$ distant receivers. After the telecloning network has been established, we can resort to either the irreversible protocol or the reversible one to realize asymmetric telecloning. In contrast to the protocols already elucidated for CV systems~\cite{Zhang_PRA_2006, Adesso_NJP_2007}, the scheme proposed here does not need additional entangled states to achieve asymmetry in the clone fidelities and is extendible to an arbitrary number of clones. 

\textit{Resource generation.} Let us first describe how the multimode entangled asymmetric resource state can be created. First, one starts with $N+1$ vacuum modes, i.e., $\ket{0}^{\otimes N+1}$. The modes are then squeezed equally in the position and momentum quadratures alternatively. In other words, mode $1$ is squeezed in the position quadrature with strength $r$, mode $2$ in the momentum quadrature with the same squeezing magnitude, and so on. Finally, the squeezed modes are combined pairwise through $N$ beam splitters, each of transmissivity $\tau_i$. Specifically, modes $1$ and $2$ are impinged on a beam splitter with transmissivity $\tau_1$, and one of the output modes is then combined with mode $3$ at a beam splitter of transmissivity $\tau_2$, and so forth. The $N+1$ output modes from the $N$ beam splitters constitute a genuinely entangled $N+1$-mode Gaussian state, characterized by $N+1$ parameters $\{r, \tau_1, \cdots, \tau_N\}$~\cite{Patra_PRA_2022}. Note that, if one considers only two modes, $\tau_1 = 1/2$ leads to the TMSV state, whereas, with three modes, the well-known Bassett-Hound state~\cite{van-Loock_PRL_2000, Adesso_NJP_2007, Adesso_JPA_2007} is obtained for $\tau_1 = 1/3$ and $\tau_2 = 1/2$. Moreover, the creation of such a resource is experimentally feasible since it involves only linear optical elements and squeezing.

\subsection{Asymmetric telecloning in the irreversible and the reversible schemes}

We assume that the mode $N+1$ belongs to the sender $\mathcal{S}$, while the other modes are distributed between the $N$ receivers. Note that, contrary to the symmetric telecloning protocols, one need not use any balanced beam splitters to create the network. Hereafter, the clones are created by following the exact steps of the irreversible and the reversible telecloning processes. The output quadratures for the clone at mode $m$ ($1 \leq m \leq N$) are given by
\begin{eqnarray}
 \nonumber \hat{x}_{m}^{out}=\hat{x}_{m}+\hat{x}_{in}-\hat{x}_{N+1},\\
    \hat{p}_{m}^{out}=\hat{p}_{m}+\hat{p}_{in}+\hat{p}_{N+1}.
    \label{eq:brau_asymm_clone_xp}
\end{eqnarray}

\noindent for the irreversible protocol, while in the reversible case,
\begin{eqnarray}
    \nonumber \hat{x}_{m}^{out}&=&\hat{x}_{m}+\hat{x}_{in}-\frac{\hat{x}_{N+1}+\hat{x}_{vS}}{\sqrt{2}},\\
\nonumber \hat{p}_{m}^{out}&=&\hat{p}_{m}+\hat{p}_{in}+\frac{\hat{p}_{N+1}+\hat{p}_{vS}}{\sqrt{2}},\\
\nonumber \hat{x}_{aC}^{out}&=&\hat{x}_{in}-\sqrt{2}\hat{x}_{vS},\\
\hat{p}_{aC}^{out}&=&-\hat{p}_{in}-\sqrt{2}\hat{p}_{vS}.
\label{eq:peng_asymm_clone_xp}
\end{eqnarray}

It is evident that the anticlone produced in the reversible protocol is resource-independent and thus has a fidelity of $1/2$ as before. In order to calculate the clone fidelity according to Eq. \eqref{eq:fidelity}, one needs the quadrature correlations corresponding to the modes $m$ and $N+1$. Analysis of the covariance matrix of an $N+1$-mode state recursively leads to the following expressions for the same.

\begin{widetext}
\begin{eqnarray}
    \nonumber \text{\bf{Variances:}} \\
    \nonumber\left<(\hat{x}_{m = 2k}^{2})\right>&=&e^{2r}-2\sinh{2r}\left\{\sum_{\substack{i=1,\\i+=2}}^{2k-1}\tau_i\prod_{j=i+1}^{2k-1}(1-\tau_j)\right\}\tau_{2k},\\
    \left<(\hat{x}_{m = 2k-1}^{2})\right>&=&e^{2r}+2\sinh{2r}\left\{1-\sum_{\substack{i=1,\\i+=2}}^{2k-3}\tau_i\prod_{j=i+1}^{2k-2}(1-\tau_j)\right\}\tau_{2k-1}.
    \label{eq:frac1}
\end{eqnarray}
\begin{eqnarray}
\nonumber \text{\bf{Correlators:}}\\
    \nonumber\left<(\hat{x}_{N+1}\hat{x}_{m=2k})\right>&=&-2\sinh{2r}\sqrt{\tau_{2k}} \sum_{\substack{i=1,\\i+=2}}^{2k-1}\tau_i\prod_{j=i+1}^{2k-1}(1-\tau_j)\prod_{l=2k}^{N}\sqrt{1-\tau_l},\\
    \left<(\hat{x}_{N+1}\hat{x}_{m=2k-1})\right>&=&\left\{1- \sum_{\substack{i=1,\\i+=2}}^{2k-3}\tau_i\prod_{j=i+1}^{2k-2}(1-\tau_j)\right\}\prod_{l=2k}^{N}\sqrt{\tau_l}2\sinh{2r}\sqrt{1-\tau_{2k-1}}.
    \label{eq:frac2}
\end{eqnarray}
\end{widetext}
While, for $N = 2k$, we have
\begin{eqnarray}
    \left<(\hat{x}_{N+1}^{2})\right>=e^{2r}-2\sinh{2r}\left\{\sum_{\substack{i=1,\\i+=2}}^{2k-1}\tau_i\prod_{j=i+1}^{2k}(1-\tau_j)\right\},
    \label{eq:frac3}
\end{eqnarray}
and for $N = 2k-1$
\begin{eqnarray}
    \left<(\hat{x}_{N+1}^{2})\right>= e^{2r}-2\sinh{2r}\left\{\sum_{\substack{i=1,\\i+=2}}^{2k-1}\tau_i\prod_{j=i+1}^{2k-1}(1-\tau_j)\right\}. 
    \label{eq:frac4}
\end{eqnarray}
Similar expressions can be found for the momentum quadratures $p_m$ and $p_{N+1}$ by performing the substitution $r\rightarrow-r$ in Eqs. \eqref{eq:frac1} - \eqref{eq:frac4}. Therefore, the fidelity of the clone produced at mode $m$ is now a function of $\{r,\tau_i's\}$. By tuning the beam splitter parameters, one can control the telecloned state at each receiver's station.

When a $5$-mode Gaussian state is distributed between a sender and four receivers, there are four beam splitters, $\tau_1, \tau_2, \tau_3,$ and $\tau_4$, involved in the resource generation process. As an exemplary state, we consider $\tau_1 = 0.5, \tau_2 = 0.05, \tau_3 = 0.125,$ and $\tau_4 = 0.1$, whence one finds that the fidelity of the clone produced at mode $1$ can easily overcome the classical threshold, but can reach $\mathcal{F}_{{C_{1}}_{\max}} = 2/3$ only for the irreversible scheme (see Fig.~\ref{fig:asymm} (a)). The clone at mode $2$ can be transmitted with quantum advantage only in a small region of the squeezing strength. However, by increasing $\tau_2$ and $\tau_3$, it can achieve the maximum fidelity at the expense of the former clone. The last two clones are always produced with a sub-classical fidelity $\mathcal{F}_{C_{3,4}} \leq 0.5$, and should thus be assigned to the untrustworthy receivers. Our measure $\mathcal{Q}_{hC_i}$, concerning the clone $i$, can again predict the fidelity accurately, with nonclassical clones being produced only when $\mathcal{Q}_{h C_i} > 0$, as is evident upon comparing Figs. $\ref{fig:asymm}$ (a) and (c) with Figs. $\ref{fig:asymm}$ (b) and (d) respectively. This scheme is easily extendable to an arbitrary number of clone modes, $N$,  thereby creating an asymmetric telecloning CV network where $\lceil N/2 \rceil$ number of receivers can obtain clones with quantum fidelity and the remaining cannot (here $\lceil z \rceil$ denotes the smallest integer greater than $z$).

\textcolor{black}{Optimal asymmetric cloning of coherent states using Gaussian maps has been analyzed previously in terms of the excess noise introduced during the protocol~\cite{Ferraro_PRA_2005,Fiurasek_PRA_2007}. 
We note that the telecloning protocol considered here is different from the one in Ref.~\cite{Ferraro_PRA_2005}, as the asymmetry in our work is characterized using transmittivity of beam splitters, without optimizing over the Gaussian resource (apart from squeezing parameter $r$), thereby leading to different optimal fidelities. 
Comparing our fidelities with the optimal conditions in Ref.~\cite{Ferraro_PRA_2005}, we observe the following: For the \textit{reversible} protocol, the fidelity of the best clone in $1\rightarrow 4$ asymmetric telecloning is equal to $2/3$, which is below the Gaussian optimal value. This is due to the fact that an anticlone is always generated in the reversible protocol with a fixed fidelity of $1/2$. On the other hand, for the \textit{irreversible} protocol, upon varying the transmittivities, the fidelity reaches very close to the Gaussian optimal value for higher values of $r$, i.e., $\mathcal{F} \approx 0.98$ at $r = 2$.}

\section{Conclusion}
\label{sec:conclu}

Telecloning entails the non-local approximate transmission of quantum information. Initially formulated for qudit systems, the protocol has been devised in the continuous variable regime, which allows for experimentally feasible realization using linear optical elements.  Through this scheme, optimal clones of a quantum state can be distributed among several spatially separated receivers with the help of multimode-entangled photons. A fundamental question concerning any quantum communication process involves understanding what specific quantum property drives the quantum advantage of the task, and also whether different classes of quantum resources can furnish different levels of advantage. In this work, we explored both these issues and established how nonclassicality present in the shared resource state can explain the telecloning technique. We studied CV telecloning in terms of two exemplary protocols - the irreversible and the reversible protocol.

We argued that neither bimodal entanglement, as quantified by logarithmic negativity, nor genuine multimode entanglement, can shed light on the optimal behavior of fidelity in a telecloning network. Entanglement is thus a necessary ingredient for telecloning although it may not be sufficient. We showed that it is the quadrature based nonclassical property, inherent in the modes shared between the sender and one of the receivers, that governs how perfectly the clones will be created. To this end, we defined a nonclassicality measure whose behavior clearly predicts the optimal fidelity of the clones. The nonclassicality measure is operationally estimable and also related to the quantumness attributed to quadrature squeezing. 

Towards identifying the best possible resource state which can provide quantum advantage, we extended the telecloning protocol by including non-Gaussian resource states. We demonstrated that two-mode states with photons subtracted from both modes can outperform the two-mode squeezed vacuum state. The advantage is ubiquitous when the reversible protocol is adopted but persists only for low resource squeezing in the irreversible scheme. The two-photon-added state can never achieve such superiority over the Gaussian TMSV resource, whereas states generated by the addition (subtraction) of photons to (from) any one mode may provide quantum advantage, albeit below that provided by the TMSV state. We also highlighted that the nonclassicality measure can explain the patterns of fidelity in the telecloning protocol for a fixed shared state. Eventually, we showed that 
only for low squeezing strengths, the fidelity can be increased when a greater number of photons is subtracted from both modes.
Our analysis thus creates an operational hierarchy among Gaussian and non-Gaussian states concerning the telecloning protocols. 

\textcolor{black}{In our work, we have focussed only on the dissipationless implementation of the telecloning
protocol. In the presence of environmental interactions, the reduction in bimodal entanglement would adversely affect the telecloning fidelities. On the other hand, it has been demonstrated that a non-Markovian nature of the interaction leads to revivals in entanglement and other non-classical resources \cite{Gupta_PRA_2022} making them reusable at several instances \cite{Nourmandipour_PRA_2016}. Hence, understanding the role of dissipation and environmental effects on the non-Gaussian resource and telecloning can further enhance the efficacy of the protocol in the CV regime. Another interesting path for future research is to find the resources for non-Gaussian advantage in protocols such as precision metrology and the capacity of quantum communication channels. It will also be fascinating to investigate whether it is possible to tweak the linear multimode teleportation protocol by introducing optical amplification (as in Ref.~\cite{Cerf_PRL_2005})
for achieving higher single-mode cloning fidelity for a subset of states.}

Going beyond the symmetric scheme, we provided an asymmetric telecloning network for continuous variable systems, where clones at different receiver stations are produced with different fidelities. It only requires linear optical elements to generate the required state, whose asymmetry with respect to different modes ensures the eventual asymmetry in the telecloning routine. Such a protocol has immense applications in scenarios that demand secure communication of information.  Our protocol allows the sender to manipulate the state parameters in order to decide which receivers would obtain clones with optimal fidelities, thereby protecting the communication scheme against malicious parties. We studied the asymmetric telecloning network under the paradigm of Gaussian states and leave its non-Gaussian implementation as an open question for further research.

\section*{Acknowledgement} 
R.G. and A.S.D. acknowledge the support from the Interdisciplinary Cyber-Physical Systems (ICPS) program of the Department of Science and Technology (DST), India, Grant No.: DST/ICPS/QuST/Theme- $1/2019/23$. This research was supported in part by the ``INFOSYS scholarship for senior students''. H.S.D. acknowledges support from SERB-DST, India under a Core-Research Grant (No: CRG/2021/008918) and
from IRCC, IIT Bombay (No: RD/0521-IRCCSH0-001).

\bibliographystyle{apsrev4-1}
\bibliography{ref}

\appendix

\section{The continuous variable phase-space formalism}
\label{app:cv}

Continuous variable (CV) systems are characterized by position, $\hat{x}$, and momentum, $\hat{p}$, operators~\cite{Serafini_2017, Braunstein_RMP_2005}. Since both the quadrature operators have an infinite spectrum, CV systems are difficult to study in the matrix notation. Therefore, one resorts to the phase-space formalism. An $N$-mode CV state, $\rho$, comprises $2N$ quadrature operators, which can be cumulatively denoted by a vector $\hat{R} = (\hat{x}_1, \hat{p}_1, \dots, \hat{x}_N, \hat{p}_N)^T$. Each quadrature operator can be written in terms of the creation and annihilation operators, $\hat{a}$ and $\hat{a}^\dagger$ respectively, as
\begin{equation}
    \hat{a}_k = \hat{x}_k + \iota \hat{p}_k, ~~~~~~~ \text{and} ~~~~~~ \hat{a}_k^\dagger = \hat{x}_k - \iota \hat{p}_k,
    \label{eq:creation-annihilation_op}
\end{equation}
where $\iota = \sqrt{-1}$ and for a given mode $k$, the commutation relation $[\hat{a}_k^\dagger, \hat{a}_k] = -1$ holds true. The commutation relation for all $N$ modes can be succinctly defined in terms of $R$ as 
\begin{equation}
    \left[\hat{R}_k,\hat{R}_l\right]= \iota \Omega_{kl}\quad \text{with} ~~  \Omega = \bigoplus\limits_{j=1}^{N} \omega_j.
  \label{eq:CV_commutation}
  \end{equation}
 Here, $\Omega$ is known as the $N$-mode symplectic form, and $\omega_j$ is given by
  \begin{equation}
      \quad \omega_j=\begin{pmatrix}
			0 & 1\\
			-1 & 0 
		\end{pmatrix} ~\forall~ j.
  \label{eq:CV_omega}
	\end{equation}
 When the Hamiltonian, $\hat{H}$, pertaining to a particular system is at most a quadratic function of the quadrature operators, one obtains the well-known \textit{Gaussian} states~\cite{ferraro2005, Weedbrook_RMP_2012} as the ground and thermal states of $\hat{H}$. As the name suggests, such states are completely specified by their first and second moments as
  \begin{eqnarray}
		&& d_k=\expval{\hat{R}_k}_{\rho}, \\
  \label{eq:CV_disp}
		\Sigma_{kl}&&= \expval{\hat{R}_k\hat{R}_l+\hat{R}_l\hat{R}_k}_{\rho}-2\expval{\hat{R}_k}_{\rho}\expval{\hat{R}_l}_{\rho},
  \label{eq:CV_cov}
	\end{eqnarray}
 where  $\mathbf{d}$ is the $2N$ dimensional \textit{displacement vector} and $\Sigma$ is the $2N \times 2N$ real symmetric and positive definite \textit{covariance matrix}. 

 Beyond Gaussian states, however, one must consider all the higher-order moments to characterize the state, which makes the analysis intractable. We can then take recourse to $m$-ordered \textit{characteristic functions}, given by~\cite{Barnett_2002, Adesso_OSID_2014}
 \begin{eqnarray}
     \chi_{\rho}^m (\mathbf{q}) = \Tr[\rho \hat{D}(\mathbf{q})] e^{m |\mathbf{q}|^2/2},
     \label{app:cv_eq1}
 \end{eqnarray}
 where $\mathbf{q}$ is the total displacement parameter of the $N$ modes, i.e., $\mathbf{q} = (q_1, q_2, \cdots, q_N)$ and $\hat{D}_k(q_k) = \exp [q_k \hat{a}_{k}^{\dagger} - q_{k}^* \hat{a}_k]$ is the displacement operator of the mode $k$ with complex displacement parameter $q_k = q_{xk} + \iota q_{pk}$. The complex Fourier transformations of $\chi_\rho(\mathbf{q}, m)$ are known as \textit{quasi-probability distributions}, $W_{\rho}^m$, i.e.,
 \begin{eqnarray}
     W_{\rho}^m (\mathbf{q}') = \frac{1}{\pi^2} \int_{\mathbb{R}^{2N}} d^{2N}\mathbf{q}''\chi_{\rho}^m(\mathbf{q}'') e^{\iota \mathbf{q}''^T \Omega \mathbf{q}'},
     \label{app:cv_eq2}
 \end{eqnarray}
 where $\mathbb{R}$ is the space of real numbers and $\mathbf{q}', \mathbf{q}'$ are displacement parameters. For $m = 0$, one recovers the Wigner function, $W_{\rho}$, which is always real, but can be both positive or negative. Thus, experimentally, one cannot directly measure the Wigner function. Its operational interpretation lies in the fact that its marginals are probability distributions that can be sampled via homodyne detection as
 \begin{eqnarray}
   \nonumber   \langle \hat{x}_k \rangle_\rho = \int_{\mathbb{R}^{2N - 1}} && dp_1 \cdots dp_N  dx_1 \cdots dx_{N-1} \\
  && W_{\rho} (x_1, p_1, \cdots, x_N, p_N).
     \label{app:cv_eq3}
 \end{eqnarray}

 \begin{widetext}
\section{Telecloning in the Wigner function formalism}
\label{app:tele_wig}

We shall discuss the telecloning protocol producing two clones of a given state (and one anticlone). The extension to $M$ clones is straightforward. Let us consider that the two-mode resource state is defined through the Wigner function $W_{\text{res}}(x_1,p_1,x_2,p_2)$ where $x_i$ and $p_i$ respectively represent the position and momentum quadratures of the mode $i$. Further, let the input state and the squeezed vacuum be characterized by $W_{\text{in}}(x_{\text{in}}, p_{\text{in}})$ and $W_v(x_v,p_v)$ respectively. To generate the telecloning network, initially, the resource state and the squeezed vacuua are impinged on balanced beam splitters. The total state may be represented as

\begin{eqnarray}
  && W_{\text{net}} =    W1_v(x_v,p_v) \times W_{\text{res}}(x_1,p_1,x_2,p_2) \times W2_v(x_v,p_v) \\
   \to && W_{\text{net}} (\frac{x_{\mathcal{S}} + x_{aC}}{\sqrt{2}}, \frac{p_{\mathcal{S}} + p_{aC}}{\sqrt{2}}, \frac{x_{C_{1}} + x_{C_{2}}}{\sqrt{2}}, \frac{p_{C_{1}} + p_{C_{2}}}{\sqrt{2}}, \frac{x_{\mathcal{S}} - x_{aC}}{\sqrt{2}}, \frac{p_{\mathcal{S}} - p_{aC}}{\sqrt{2}}, \frac{x_{C_{1}} - x_{C_{2}}}{\sqrt{2}}, \frac{p_{C_{1}} - p_{C_{2}}}{\sqrt{2}}), \label{eq:app1}
\end{eqnarray}
where $\mathcal{S}$ denotes the mode at the sender, $C_i$ denotes the clone $i$, and $aC$ denotes the anticlone mode. Note that in the irreversible protocol, the state $W1_v$ is absent and no anticlones are produced.

Upon establishing the setup, the sender combines the input state and his/her mode at a balanced beam splitter to undertake homodyne detection.
\begin{eqnarray}
    && W' = W_{\text{in}} (x_{\text{in}}, p_{\text{in}}) \times W_{\text{net}}(x_\mathcal{S}, p_\mathcal{S}, x_{aC}, p_{aC}, x_{C_{1}}, p_{C_{1}}, x_{C_{2}}, p_{C_{2}}) \\
    \to && W'(\frac{x_u + x_w}{\sqrt{2}}, \frac{p_u + p_w}{\sqrt{2}}, \frac{x_w - x_u}{\sqrt{2}}, \frac{p_w + p_u}{\sqrt{2}}, x_{aC}, p_{aC}, x_{C_{1}}, p_{C_{1}}, x_{C_{2}}, p_{C_{2}}), \label{eq:app2}
\end{eqnarray}
where $u$ and $w$ represent the output modes from the beam splitter. The sender performs homodyne detection on the modes $x_u$ and $p_w$, which translates to integrating Eq. \eqref{eq:app2} over the corresponding variables. To facilitate calculations, we substitute $x_u + x_w \to \sqrt{2} x, p_u + p_w \to \sqrt{2} p$ and obtain $W'(x_u, p_w, x, p, x_{aC}, p_{aC}, x_{C_{1}}, p_{C_{1}}, x_{C_{2}}, p_{C_{2}})$. The protocol specifies that the anticlone mode be displaced by $-\sqrt{2}x_u + \iota \sqrt{2} p_w$ and each clone mode be displaced by $-\sqrt{2}x_u - \iota \sqrt{2} p_w$ (here $\iota = \sqrt{-1}$). Therefore,
\begin{equation}
    W' \to W(x_u, p_w, x, p, x_{aC}-\sqrt{2}x_u, p_{aC} +  \sqrt{2} p_w, x_{C_{1}}-\sqrt{2}x_u, p_{C_{1}} - \sqrt{2} p_w, x_{C_{2}}-\sqrt{2}x_u, p_{C_{2}} - \sqrt{2} p_w). \label{eq:app3}
\end{equation}

Then the Wigner function of any clone, say $C_1$, can be readily obtained as $W_{C_{1}} (x_{C_{1}}, p_{C_1})$ by integrating Eq. \eqref{eq:app3} over $x_u, p_w, x, p, x_{aC}, p_{aC}, x_{C_{2}}, p_{C_{2}}$. The fidelity of the clone is then given by
\textcolor{black}{
\begin{eqnarray}
   \nonumber \mathcal{F}_{C_{1}} && = 2\pi\int dx_{C_{1}} dp_{C_{1}} dx_{\text{in}} dp_{\text{in}} W_{C_{1}} (x_{C_{1}}, p_{C_1}) \times W_{\text{in}}(x_{\text{in}}, p_{\text{in}}) \delta(x_{C_{1}} - x_{\text{in}}) \delta(p_{C_{1}} - p_{\text{in}}) \\
   && = 2\pi \int  dx_{\text{in}} dp_{\text{in}} W_{C_{1}} (x_{\text{in}}, p_{\text{in}}) \times W_{\text{in}}(x_{\text{in}}, p_{\text{in}}).
   \label{eq:app4}
\end{eqnarray}
}

Note that the Wigner function of any other clone, or the anticlone, can be obtained by integrating Eq. \eqref{eq:app3} over all the remaining mode quadratues. For a complicated network of $1 \to M$ telecloning, more balanced beam splitters are used to split one resource mode into the $M$ modes in Eq. \eqref{eq:app1}. \textcolor{black}{In our work, all integrations involving non-Gaussian states have been performed numerically. }

\end{widetext}

\end{document}